\documentclass[prd,a4paper,nofootinbib,
showpacs, 
]{revtex4}
\usepackage{graphicx}
\def\eq#1{{eq.~(\ref{#1})}}
\def\eqs#1#2{{eqs.~(\ref{#1})--(\ref{#2})}}

\def\vev#1{\left\langle #1\right\rangle}

\def\hbar{\hspace{0pt}\raisebox{1pt}{$-$} \hspace{-7pt} h}

\def\5{\overline 5}

\newcommand{\be}{\begin{equation}}
\newcommand{\ee}{\end{equation}}
\newcommand{\bea}{\begin{eqnarray}}
\newcommand{\eea}{\end{eqnarray}}
\newcommand{\nn}{\nonumber}

\begin{document}
\title[Just so Higgs boson ]{Just so Higgs boson}
\date{\today
}
\author{F.~Bazzocchi}
\affiliation{IFIC - Instituto de F\'\i sica Corpuscular\\
University of Valencia, Apartado de Correos 22085 \\ 
E-46071 Valencia, Spain}
\author{M.~Fabbrichesi}
\author{P.~Ullio}
\affiliation{INFN, Sezione di Trieste and\\
Scuola Internazionale Superiore di Studi Avanzati\\
via Beirut 4, I-34014 Trieste, Italy}
\begin{abstract}
We discuss a minimal extension to the standard model in which there are two Higgs bosons  and, in addition to the usual fermion content, two  fermion doublets and one fermion singlet. The little hierachy problem is solved by the vanishing of the one-loop corrections to the quadratic terms of the scalar potential. The  electro-weak ground state is therefore stable for values of the cut off up to 10 TeV. The Higgs boson mass can take values significantly larger than the current LEP bounds and  still be consistent with electro-weak precision measurements. 
\noindent  
\end{abstract}
\maketitle
%
\vskip1.5em
\section{Motivations} 
\label{sec:mot}
There is some  tension between the  value of the electroweak (EW) vacuum and the scale at which we expect new physics to become manifest according to EW precision measurements~\cite{PDG}. If we take the latter scale around 10 TeV as the cutoff of our effective theory,  some degree of fine tuning is necessary in the scalar potential in order to guarantee  the vacuum stability  against radiative corrections. This little hierachy problem---and before it the more general (and more serious) problem of  the large hierarchy between the EW vacuum and the GUT and Planck scale---has been used as a clue to the development of  models in which  the scalar sector of the standard model is enlarged to provide better stability, as, for instance, in supersymmetry, technicolor and little-Higgs models. 

Here we discuss a different approach in which no new symmetry is introduced to cancel loop corrections and instead the parameters of the  lagrangian are  such as to make the one-loop corrections vanish and thus ensure the stability of the effective potential for the scalar particles up to the energy scale at which two-loop effects begin to be sizable, namely 10 TeV. Clearly, by its very nature, such a procedure  can only be applied to the little hierarchy problem and not to the more general GUT or Planck scale hierarchy problem. 
 It is a limited solution to  a little (hierarchy) problem, a problem that---contrary to those arising in much larger hierarchies---may well be contingent to the choice of the lagrangian parameters. 
 
 Given the simplicity of the idea behind this approach, it is not surprising that it was suggested early on (by Veltman~\cite{veltman}) in the following terms: the quadratically divergent one-loop correction to the Higgs boson mass $m_h$,
 \be
 \frac{3 \Lambda^2}{16 \pi^2 v_W^2} \left[ 2 m_W^2 + m_Z^2 +m_h^2 - 4 m_t^2 \right] \, , \label{velt1}
\ee
can be made to vanish, or at least made small enough, if  $m_h$ happens to be around 316 GeV at the tree level. The remaining contributions---not included in (\ref{velt1})---are proportional to the light quark masses and therefore negligible.

Such a cancellation does not originate from any dynamics and it is the accidental result of the values of the physical parameters of the theory. The absence of this quadratically divergent term in the two-point function of the scalar bosons makes possible to increase the  cut-off for the theory to a higher value with respect to the standard model (SM) where the renormalization of the Higgs boson mass and the given value of the expectation value $v_W$ impose a cut off of around  1 TeV to avoid unnaturally precise cancellations among terms.

We now know that a value of $m_h = 316$ GeV is a little over   3$\sigma$ with respect to  current precision measurements of EW data~\cite{PDG}. This however does not mean that a scenario in which the Higgs boson mass is chosen just so to make the cancellation \textit{\`a la} Veltman is ruled out. It only means that we must either enlarge the SM with new particles propagating below 1 TeV and then redo the EW data fit~\cite{Peskin0} or introduce new physics at a higher scale, the effect of which is to correct the precision observables and make room for the shifted value of the Higgs boson mass (as described in the framework of the effective EW lagrangian in~\cite{Bagger}). 

Bearing this in mind, we introduce the minimal extension to the SM in which
\begin{itemize}
\item quadratically divergent contributions cancel at one-loop  \textit{\`a la} Veltman;
\item it is consistent with the EW precision data.
\end{itemize}
The model, as we shall see, is quite simple and 
provides an explicit example of an extension of the SM in which the mass of the Higgs boson can assume significantly larger values with respect to the current lower bound without having the EW precision measurements violated. In so doing, it introduces a characteristic spectrum of states beyond the SM that can be investigated at the LHC.

The model is natural in the sense that the EW vacuum is stable against a cut off of the order of 10 TeV for a large choice of parameters. It is just so because the physical parameters are chosen by hand in order to satisfy the constraints. These parameters are however numbers of order unity and not extravagantly small or large; moreover, they can be chosen among many possible values so that no unique determination is required, as it would be in the original Veltman's condition where the only free parameter is the Higgs boson mass, or---what amounts to the same thing---the quartic coupling $\lambda$. 

Because among the additional states required there is a stable neutral (exotic) fermion, we also discuss to what extent this state can be  considered a candidate for  dark matter. 

\vskip1.5em
\section{The model: How the Higgs got its mass} 
\label{sec:model}
We consider a model in which there are two scalar EW doublets, $h_1$ and $h_2$, the lightest scalar component of which is going to be identified with the SM Higgs boson and two Weyl fermion doublets, $\psi_1$ and $\psi_2$. In addition, we also introduce a Weyl fermion $\psi_3$ which is a $SU(2)_W\times U(1)_Y$ singlet.

Let us briefly discuss to what extent this choice of new states is the minimal extension which cancels the quadratic divergence. In the model with only two Higgs bosons it is possible to reduce---or indeed cancel---the contribution of the top quark to the quadratic divergence but not that of the gauge bosons and the cutoff cannot be raised up to 10 TeV. We comment on this class of models in sec.~\ref{sec:2} below. 
The mass of  a single fermion doublet (with two singlets) is necessarily proportional to the scalar field vacuum expectation and cannot be varied independently of the Veltman condition (if we want to choose naturally the Yukawa of the new fermions). Two doublet fermions are also necessary in order to be anomaly free. The singlet fermion is necessary to lift the fermion degeneracy and couple the fermions to the scalar fields.

The states of the model are similar to those of a SUSY minimal extension of the scalar sector of the SM into a Wess-Zumino chiral model in which the singlet boson has been integrated out. However, a model with  softly broken supersymmetry cannot be the model we are discussing because the supersymmetry, if present at any scale, would make the quadratically divergence zero. 

The lagrangian for the scalar bosons is given by
\be
{\cal L}_h = \sum_{i=1,2} D_\mu h^i D^\mu h^i + V[h_1,\ h_2]
\ee
with the potential 
\bea
V[h_1,\ h_2] & = &\lambda_1 (h_1^\dag h_1)^2 + \lambda_2 (h_2^\dag h_2)^2 + \lambda_3 (h_1^\dag h_1) (h_2^\dag h_2) + \lambda_4 (\tilde h_2^\dag h_1) (\tilde h_1^\dag h_2) + \lambda_5 \left[(\tilde h_2^\dag h_1)^2 + H.c. \right] \nn \\
&+ &  \mu_1^2 (h_1^\dag h_1) + \mu_2^2 (h_2^\dag h_2) \, ,
\label{potB}
\eea
where $\tilde h_2 =i \sigma_2 h_2^* $. 
The potential in \eq{potB} is the most general for the two Higgs doublets once we impose a parity symmetry $T_1$ according to which the two doublets $h_1$ and $h_2$ are, respectively even and odd. In this way, the quadratic and quartic mixing terms are forbidden, which makes the discussion simpler.
 
 The lagrangian of \eq{potB} can be studied  to find the ground state that triggers the electroweak symmetry breaking. It is
\be
\label{vevs}
\vev{h_1} = {0 \choose v_1/\sqrt{2}} \quad \mbox{and} \quad \vev{h_2} = {v_2/\sqrt{2} \choose 0}\, .
\ee
The requirement of matching the EW vacuum $v_w$ to this vacuum state constains one parameter of the model. 

The mass eigenstates of the scalar particles can thus be derived. 
The masses are
\bea
\label{masseneutri}
m_{h,H}^2 & =&   \lambda_1 v_1^2 +  \lambda_2 v_2^2 \pm \sqrt{ \left(\lambda_1 v_1^2 - \lambda_2 v_2^2 \right)^2 +  \left( \lambda_3 + \lambda_4 + \lambda_5 \right)^2 v_1^2v_2^2} \\\nn
m_A^2 &= & - \lambda_5 v_w^2\,,
\eea
for the three neutral scalar bosons (two of which, $h$ and $H$, are scalars  and one, $A$,  a pseudoscalar),
\be
\label{massacarico}
m_{H^{+}}^2 = - \left( \lambda_4 + \lambda_5 \right) v_w^2\,,
\ee
for the charged boson $H^+$ after using the constraint $v_w^2=\left(v_1^2 + v_2^2 \right) $  in \eqs{masseneutri}{massacarico}.

By introducing the  mixing angle $\alpha$ and $\beta$ to rotate the scalar boson gauge states into the mass eigenstates, we write:
\be
\label{hruot}
h_1 =  \frac{1}{\sqrt{2}} {\sin\beta\, H^+ \choose v_1 + \cos\alpha\, h -\sin \alpha\, H+ \sin\beta\,A} \quad \mbox{and} \quad 
h_2 = \frac{1}{\sqrt{2}}{v_2 + \sin \alpha\, h+\cos \alpha\, H + \cos \beta\, A \choose \cos \beta \,H^-}\, .
\ee
 As usual in $2$ Higgs doublet model $\tan\beta =v_2/v_1$ and 
\be
  \label{eqalfa}
\tan 2\alpha =
\frac{\lambda_3 v_1 v_2}{(\lambda_2 v_2^2-\lambda_1v_1^2)}\, .
\ee
 
The exotic fermion content of the model is given by  two  $SU(2)_W$ doublets:
 \bea
{\Psi}_1=\, \left (\begin{array}{c}
  \psi_1^{+} \\
\psi_1^{0}
\end{array}\right)\,\quad& {\Psi}_2=\, \left (\begin{array}{c}
  \psi_2^{0} \\
\psi_2^{-}
\end{array}\right) \nn\,,
\eea
and one $SU(2)_W$ singlet $\psi_3$;  we can also define the Majorana 4-components  fermions current eigenstates as
 \bea
 \label{majint}
\tilde{\psi}_i^{0}=\, \left (\begin{array}{c}
  \psi_i^{0} \\
\bar{\psi}_i^{0}
\end{array}\right) &\quad & \tilde{\chi}_i^{+}=\, \left (\begin{array}{c}
  \psi_1^{+} \\
\bar{\psi}_2^{-}
\end{array}\right)
 \nn\,.
\eea

The SM fermions are even under the $T_1$ parity symmetry and therefore can have Yukawa interactions only with the scalar doublet $h_1$. The exotic doublet fermions $\Psi_i$  are odd under this parity symmetry while the singlet $\psi_3$ is even. We also introduce an additional parity $T_2$ under which the Higgs bosons are even while all the exotic fermions are odd (SM particles are always even under both parities). In this way the exotic fermions do not mix with the SM fermions and may have Yukawa terms only with the scalar doublet $h_2$.

 The lagrangian for the exotic fermions is simply given by the kinetic and the Yukawa terms, that is
 \be
 \mathcal{L}^{\psi}=\mathcal{L}_{kin}+\mathcal{L}_m^\psi \,,
 \ee
where $\mathcal{L}_{kin}$ is given by 
\bea
\label{Lkin}
\mathcal{L}_{kin}&=& \bar{\Psi}_1 \hat{D}  \Psi_1 +\bar{\Psi}_2 \hat{D}  \Psi_2 + \bar{\psi_3}\hat{\partial} \psi_3\nn
\eea
and
\bea 
\label{Lm}
-\mathcal{L}_{m}&=&\tilde{\mu} \,\epsilon_{ij}\Psi_{1_{i}}\Psi_{2_{j}}+ \hat{\mu}\,\psi_3 \psi_3 +\left( \frac{k_1}{\sqrt{2}} \psi_3 \tilde{h_2}^\dag \Psi_1 + H.c. \right) \nn\\
&+& \left( \frac{ k_4}{\sqrt{2}} \psi_3 \Psi_{2_{i}} \tilde{h_{2}}_{j}  + H.c. \right) \nn\,.
\eea
From $\mathcal{L}_m$ we see that the charged Dirac  fermion $\tilde{\chi}^{+}$
has mass $m_{\chi^+}= \tilde{\mu}$ while once we insert \eq{vevs} into \eq{Lm} the Majorana mass matrix for the neutral states is given by
 \bea
 \label{majneut}
 M^0=\left( \begin{array}{ccc}
 0& -\tilde{\mu}& k_1 v_2/\sqrt{2}\\
 -\tilde{\mu}&0&k_4 v_2/\sqrt{2}\\
 k_1 v_2/\sqrt{2}&k_4 v_2/\sqrt{2}& \hat{\mu} \,.
 \end{array}\right)
 \eea
This matrix is diagonalized by a neutralino mixing matrix $V$ which satisfies $V^{T} M^0 V^*= M^0_{diag}$. From the 2-components mass eigenvectors of \eq{majneut} $\chi^0_{i=1,3}$, we define the  4-components neutral fermions that will be our neutralinos
 \bea
 \label{eigmass}
\tilde{\chi}_1^0= \left (\begin{array}{c}
  \chi_1^{0} \\
\bar{\chi}_1^{0}
\end{array}\right)\nn &  \tilde{\chi}_2^0=\gamma_5 \left (\begin{array}{c}
  \chi_2^{0} \\
\bar{\chi}_2^{0}
\end{array}\right)\nn&\tilde{\chi}_3^0= \left (\begin{array}{c}
  \chi_3^{0} \\
\bar{\chi}_3^{0}
\end{array}\right)\nn \,,
\eea
 where in \eq{eigmass} the definition of $\tilde{\chi}_2^0$ takes into account that the corresponding eigenvalue of the Majorana  mass matrix is negative.

From $\mathcal{L}_{kin}$ of \eq{Lkin}
using the mass eigenstates defined in \eq{eigmass} we obtain the interaction terms of the new fermions with the gauge bosons
\bea 
\mathcal{L}&=& \frac{g}{\sqrt{2}} \Big [\sum_{i=1,3}
\bar{\tilde{\chi}}^{+} \gamma_\mu (V_{1\,i}\,\epsilon_i P_L-
V^*_{2\,i}P_R) \tilde{\chi_i}^{0}\Big] W^{+\,\mu}+
H.c.+ \frac{g}{2} \bar{\tilde{\chi}}^{+} \gamma_\mu (P_L+P_R)
\tilde{\chi}^{+}W_3^\mu \nn\\
& - & \frac{g}{2} \sum_{i,j=1,3}\Big[ \bar{\tilde{\chi_i}}^{0}
\gamma_\mu \big(\epsilon_i\,
\epsilon_j(V^\dag_{i\,1}V_{1\,j}-V^\dag_{i\,2}V_{2\,j}) P_L-P_R 
(V^T_{i\,1}V^*_{1\,j}-V^T_{i\,2}V^*_{2\,j})\big)
\tilde{\chi_j}^{0}\Big]W_3^\mu \nn\\
&+&
\frac{{g'}}{2}\bar{\tilde{\chi}}^{+} \gamma_\mu (P_L+P_R)
\tilde{\chi}^{+}B^\mu \\
& + & \frac{{g'}}{2} \sum_{i,j=1,3}\Big[ \bar{\tilde{\chi_i}}^{0}
\gamma_\mu \big(\epsilon_i\,
\epsilon_j(V^\dag_{i\,1}V_{1\,j}-V^\dag_{i\,2}V_{2\,j}) P_L-P_R 
(V^T_{i\,1}V^*_{1\,j}-V^T_{i\,2}V^*_{2\,j})\big)
\tilde{\chi_j}^{0}
  \Big]B^\mu \nn\,,
 \eea 
  where  the factor $\epsilon_i=(-1)^{i-1}$ keeps
into account the signs of the eigenvalues of the Majorana mass matrix of \eq{majneut}. This lagrangian is necessary in order to compute the one-loop radiative corrections to the scalar potential.
 
\vskip1.5em
\section{Veltman condition redux} 
\label{sec:velt}
As stated in the introduction, we want to stabilize the potential given by \eq{potB} at one-loop level, that is we want that the one-loop $\delta \mu_i^2$ quadratically divergent contributions to  $\mu_i^2$ be zero.  As in the SM the quadratically divergent contributions arise by loops of gauge bosons, scalars and fermions.  We therefore find  two Veltman conditions  by imposing
\be
\delta \mu^2_1=0\quad \mbox{and} \quad \delta \mu^2_2=0\,,
\ee
that is
\bea
\label{velt}
\frac{9}{4} g^2 + \frac{3}{4} {g'}^2  +2(3\lambda_1+\lambda_3+\lambda_4)  -12 \lambda_t^2 & =& 0 \nn \\
\frac{9}{4} g^2 + \frac{3}{4} {g'}^2    +2(3\lambda_2+\lambda_3+\lambda_4)   -(k_1^2+k_4^2)    & =& 0\,.
\eea
In \eq{velt}  $g$, ${g'}$ are the electroweak gauge couplings, $\lambda_i$ the parameter of the scalar potential of \eq{potB} , $\lambda_t$  the top Yukawa defined as $\lambda_t=v_w/v_1$ since the SM fermions couple only to the scalar doublet $h_1$ and $k_{1,4}$ are the Yukawa coupling of \eq{Lm}. The contributions of the lighter SM fermions to \eq{velt} have been neglected.

Notice that if we did not have the parity symmetry $T_1$ and the fermions would have interacted with both  $h_1$ and $h_2$ we would have generated a divergent mixed  contribution that could have been canceled only by a bare term.

In writing \eq{velt} we have taken a common cutoff $\Lambda$ for the divergent loops of different states. The possibility that there exist different cutoffs for the different contributions does not change our result because a change of order $O(1)$ in the $\Lambda$s only means a similar change of order $O(1)$ in the parameters of the model $\lambda_i$ and $k_i$. 
 
Once these two conditions are satisfied the scalar potential is stable at one-loop order and so is its vacuum state. We interpret these conditions as two constraints on the 10 free parameters of the model.

\vskip1.5em
\section{The EW parameters $S$, $T$ and $U$} 
\label{sec:stu}
For our purposes, the consistence of the model against EW precision measurements can be checked by means of oblique corrections.
These corrections can be classified~\cite{Peskin} by means of three parameters:
\bea
\alpha S & = &- 4 e^2 \left[\Pi_{33}'(0) - \Pi_{3B}' (0) \right] \nn \\
\alpha T & = & \frac{e^2}{s_W^2 c_W^2 m_Z^2} \left[ \Pi_{11}(0)- \Pi_{33} (0) \right] \nn \\
\alpha U &=& 4 e^2 \left[ \Pi_{11}' (0)- \Pi_{33}' (0) \right] \, ,
\eea
where the functions $\Pi_{nm}(q^2=0)$ represent the vacuum polarizations of the gauge vactors in the various directions of  isospin space. Other corrections functions---like the functions $Y$ and $W$ of ref.~\cite{strumia}---are not relevant here because mainly sensitive to physics in which there are new vector bosons.

EW precision measurements severely constrain the possible values of the three parameters S, T and U. In the SM, the data allow~\cite{PDG}, for a Higgs boson mass of $m_h=$ 117 GeV,
\bea
S &=& -0.13  \pm 0.10 \nn \\
T &=& -0.17   \pm 0.12 \nn \\
U &= & 0.22\pm   0.13 
\eea
These constraints must be rescaled for the different values of the Higgs boson mass. If we want the model to be consistent with the EW precision measurements within, for instance, one sigma  we have three further constraints on the parameters of the model---5 of which still remain free at this point. 

A mass of the Higgs boson larger than the reference value  will make the parameter $T$ smaller, the size of the correction going like the $\ln m_h^2/m_Z^2$. This can be compensated by the fermion contribution which can give a $\Delta T >0$ of size $\Delta m^2 \ln m^2$ where $\Delta m^2$ is the isospin splitting of the fermion masses. The parameter $S$ is changed by the larger Higgs mass with a $\Delta S>0$, a change that is in general difficult to compensate. In our model a negative contribution to $S$ comes about because of the fermion with both Dirac and Majorana masses which gives a negative contribution proportional to $\ln (m_{\chi^{+}}/m_{\chi^{0}_{i}})$, and $m_{\chi^{+}}-m_{\chi^{0}_{i}}$ is the  isospin mass splitting between the chargino and the neutralino $i$.

 Let us consider their contributions to $T$ and $S$ separately in a simplified model which helps in  visualizing better how scalar and  fermion contributions compensate one other in order to  accomodate the EW experimental values. 

For what concerns the fermions, suppose to be in the simple case in which $m_{\chi^{+}}\sim  m_{\chi^{0}_{i}}$. The fermion contribution to the $T$ parameter   can be written as
\be
T^f=T^f_{LL}+T^f_{LR}\, ,
\ee
where $T^f_{LL}$ takes into account the one loop contributions that arise from the vacuum polarizations of the gauge bosons of the kind $\Pi^{LL}_{11,33}$ and $\Pi^{RR}_{11,33}$,  $T^f_{LR}$ the ones that arise from  $\Pi^{LR}_{11,33}$ and $\Pi^{RL}_{11,33}$. The latters  are not present in the SM case.  Keeping only the leading contributions, we have
\bea
T^f_{LL} &=& \frac{2}{c_W^2 s_W^2 m_Z^2 \pi}
\Big\{ \sum_i U^{eff}_i \Big[-m^{2}_{\chi^{+}} +\frac{( m_{\chi^{+}}-m_{\chi_{i}} )^2}{2} \Big] \log \frac{m^{2}_{\chi^{+}}}{m_Z^2} \nn\\
&-&\sum_{i,j} \frac{U^{eff}_{ij}}{2} \Big[ -m^{2}_{\chi^{0}_{i}} +\frac{( m_{\chi^{0}_{i}}-m_{\chi^{0}_{j}} )^2}{2}\Big] \log \frac{m^{2}_{\chi^{0}_{i}}}{m_Z^2}\Big\} \nn\\
T^f_{LR}&=& \frac{2}{c_W^2 s_W^2 m_Z^2 \pi}\Big\{ \sum_i U^{eff}_i m_{\chi^{+}}m_{\chi^{0}_i} \log \frac{m^{2}_{\chi^{+}}}{m_Z^2} \nn\\
&-&\sum_{i,j} \frac{U^{eff}_{ij}}{2}   m_{\chi^{0}_{j}}m_{\chi^{0}_{i}}  \log \frac{m^{2}_{\chi^{0}_{i}}}{m_Z^2}\Big\}\,,
\eea
where $U^{eff}_{i,ij}$ are effective couplings related to the neutralino mixing matrix $V$ and are in general different for the $LL$ contribution and for the $LR$ one. For example  $U^{eff}_i$ in $T^f_{LL}$ is given by $\Sigma_k=V^{\dag}_{ik}V_{ki}$.
Notice that when the third neutralino decouples, the $T^f_{LL}$ contribution goes to zero when $m_{\chi^{+}} =m_{\chi^{0}_{1,2}} $  and the isospin symmetry is restored. The same  happens also for $T^f_{LR}$.
In a manner similar to the $T$ parameter, the $S$ parameter receives a fermion contribution that can be splitted in
\be
S^f= S^f_{LL}+S^f_{LR}\,,
\ee
with
\bea
S^f_{LL}&=&  \frac{1}{3 \pi}\left[(U_{ij}^{eff}-1)-2 \log \frac{m_{\chi^{+}}^2}{m_Z^2} + 2 U_{ij}^{eff} \log \frac{m_{\chi^{0}_{i}}^2}{m_Z^2}\right] \nn\\
S^f_{LR}&=& \frac{2}{3 \pi} (y_{R_{\chi^{+}}} -  y^{eff}_{R_{\chi^{0}_{ij}}})\,,
\eea
where $y_{R_{\chi^{+}}}=1/2$ follows by the definition of $\chi^{+}$ and where we have defined  $y^{eff}_{R_{\chi^{0}_{ij}}}=  2 \Sigma_{k=1,2} T_{3_{k}} y_{R_{k}} f(V_{ki},V_{kj})$ where $T_{3_{k}} $and  $y_{R_{k}} $ are the isospin and the hypercharge of the Majorana singlets defined in \eq{majint} and $ f(V_{ki},V_{kj})$ is a combination of different entries of the neutralino mixing matrix $V$.  Notice that we recover the contribution of two   SM-like doublets when the hypercharges difference in $S^f_{LR}$ gives $1/2$ and  $U_{ij}$  in $S^f_{LL}$ is equal to $1$\cite{Peskin}. 

\begin{figure}
\begin{center}
\includegraphics[width=5in]{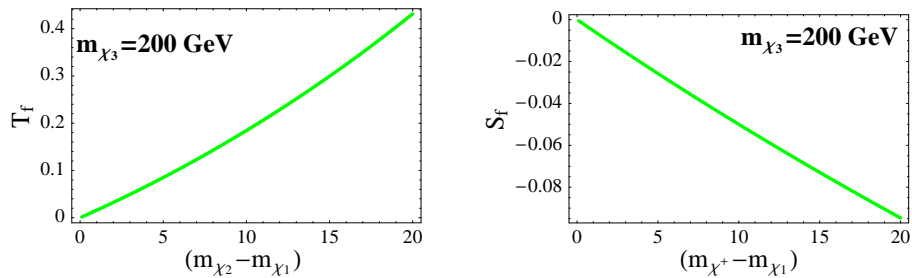}
\caption{\small  Fermion contributions to the parameters $T$ and $S$ as a function of their mass splitting. The plots are made for one representative value of the $\chi^0_3$ mass. \label{fig1}}
\end{center}
\end{figure}

The previous expressions can be further simplified if we  consider the fermion mass matrix of \eq{majneut} in the limit in which $\hat{\mu}$ is much larger than $\tilde{\mu},k_{1,4} v_2$. In this limit the neutral fermion mixing matrix is approximately given by 
\bea
\label{Ufapprox}
V &=& \left(\begin{array}{ccc} -\frac{1}{\sqrt{2}}&  -\frac{1}{\sqrt{2}}& \sqrt{\frac{m_{\chi^{0}_{2}}-m_{\chi^{0}_{1}}}{ 2 m_{\chi^{0}_{3}}}}\\   \frac{3 m_{\chi^{0}_{1}}-m_{\chi^{0}_{2}}}{2 \sqrt{2}m_{\chi^{0}_{1}} } &-\frac{m_{\chi^{0}_{1}}+m_{\chi^{0}_{2}}}{2 \sqrt{2}m_{\chi^{0}_{1}} } & -\frac{m_{\chi^{0}_{3}} }{m_{\chi^{0}_{1}}}\sqrt{\frac{m_{\chi^{0}_{2}}-m_{\chi^{0}_{1}}}{ 2 m_{\chi^{0}_{3}}}}\\\
\frac{m_{\chi^{0}_{1}}+m_{\chi^{0}_{3}}}{m_{\chi^{0}_{1}}}\sqrt{\frac{m_{\chi^{0}_{2}}-m_{\chi^{0}_{1}}}{ 2 m_{\chi^{0}_{3}}}} & \frac{m_{\chi^{0}_{1}}-m_{\chi^{0}_{3}}}{m_{\chi^{0}_{1}}} \sqrt{\frac{m_{\chi^{0}_{2}}-m_{\chi^{0}_{1}}}{ 2 m_{\chi^{0}_{3}}}}&1 \end{array}   \right)\,.
\eea
If it holds also that  $\tilde{\mu}$ larger than $k_{1,4} v_2$,  $m_{\chi^{+}}\simeq m_{\chi^{0}_{1,2}} $, $T^{f}$ and $S^f$ can be easily expressed in terms of the mass splitting, see fig.~\ref{fig1}. Notice that \eq{Ufapprox} is valid  for $m_{\chi^{0}_{1,2}}< m_{\chi^{0}_{3}}< m^2_{\chi^{0}_{1}}/\Delta m_{\chi^{0}_{12}}$. For this reason we have plotted $T^f$ and $S^f$ in fig.~\ref{fig1} corresponding to only one value of $m_{\chi^{0}_{3}}$, since in the range allowed  differences are minimal.

\begin{figure}
\begin{center}
\includegraphics[width=5in]{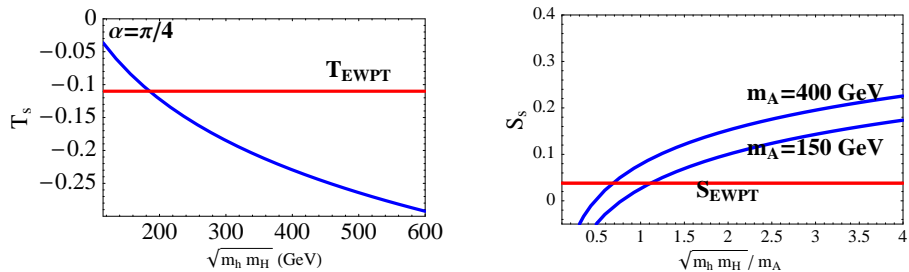}
\caption{\small   Scalar contributions to the parameters $T$ and $S$ as a function of their masses. The red horizontal lines show the central value of the current EW bounds. Notation and values of the parameters are explained in the text. \label{fig2}}
\end{center}
\end{figure}

For what concerns  the scalar sector, consider the case in which $m_{H^+}\simeq m_A$. In this limit, $T^s$ and $S^s$ assume a  simple form. We have 
\be
T^s= -\frac{3}{16 \pi c_W^2}(\cos^2 \alpha \log \frac{m^2_h}{m_Z} +\sin^2 \alpha \log \frac{m^2_H}{m_Z})\,,
\ee
where $ \alpha$ is the mixing angle in the neutral scalar sector and
\be
S^s= \frac{1}{12 \pi}( \log \frac{m^2_h}{m^2_Z} + \log\frac{m^2_H}{m^2_Z}-2 \log \frac{m^2_{H^{+}}}{m^2_Z}+\log\frac{m^2_A}{m^2_Z})\,.
\ee
We can compare the contribution of the scalar sector of our model with  respect to the SM one.  In the SM we have
\bea
T^{SM}_{h}&=& -\frac{3 }{8 \pi c_W^2} \log \frac{m_h}{m_Z}\nn\\
S^{SM}_{h}&=& \frac{1}{12 \pi}\log \frac{m^2_h}{m^2_Z}\,.
\eea
EW precision measurements indicate that at $2\sigma$ $m_h\leq 185$ GeV~\cite{PDG} and therefore the introduction of the fermions in our model is justified if $T^s$ and $S^s$   exceeds the contributions $T^{SM}_h$ and $S^{SM}_h$  corresponding to $m_h \simeq 185$ GeV. This is shown in fig.~\ref{fig2}.

For fixed $m_h$ and $m_H$, and a given  fermion spectrum that accomodates the $T$ parameter,  the fermion contribution to $S$ is fixed and therefore the only freedom left is in the values of $m_A$ and $m_{H^{+}}$. In the case in which $m_{H^+}\simeq m_A$, their total contribution to $S$ has the same sign of the fermion one,  therefore we expect that $m_A$ cannot in general be to heavy. This is verified in the numerical analysis.

\vskip1.5em
\section{A dark matter candidate?} 
\label{sec:dm}
The lightest neutral exotic fermion state in the model is  similar to the neutralinos in a minimal supersymmetric extension of the SM (NMSSM) in which the composition is dominated by Higgsinos. It is stable because the lagrangian  does not contain couplings between  the SM and the exotic fermions---or, alternatively, you can think of the lagrangian as written with a underlying  conserved parity. 

We  compute by means of the program \texttt{DARKSUSY}~\cite{DS}
its relic abundance $\Omega_{DM} h^2$.  To do this we need the lagrangian written on the exotic fermion mass eigenstates:
\bea 
-2 \mathcal{L}_m&=& \sum_{i=1,2,3}\,{m_i}
\bar{\tilde{\chi}}_i^{0} \chi^0_i+ \sum_{i,j=1,2,3}
\bar{\tilde{\chi}}_i^{0} \, (V^T_{i\,3} V_{1\,j} \epsilon_j P_L+ V^\dag_{i\,3} V^*_{1\,j} \epsilon_i P_R)\tilde{\chi}^{0}_j\nn\\
&&[\sum_{n=1,2}\,(k_1 U^R_{2\,n}+k_2 U^R_{1\,n})H_n]\nn\\
&+& \sum_{i,j=1,2,3}\,\bar{\tilde{\chi}}_i^{0}  \,
(V^T_{i\,3} V_{2\,j} \epsilon_j P_L+V^\dag_{i\,3} V^*_{2\,j} \epsilon_i
P_R)\tilde{\chi}^{0}_j [\sum_{n=1,2}(k_4 U^R_{2\,n}+ k_3
U^R_{1\,n})H_n]\nn
\\
&+&\sum_{i,j=1,2,3}\,
\bar{\tilde{\chi}}_i^{0} \, [(-i \,V^T_{i\,3} V_{1\,j} \epsilon_j P_L+ i \,V^\dag_{i\,3} V^*_{1\,j} \epsilon_i P_R)(k_1
\cos\beta+k_2 \sin\beta) \nn\\
&+& \, (i \,V^T_{i\,3} V_{2\,j} \epsilon_j P_L- i \,V^\dag_{i\,3} V^*_{2\,j} \epsilon_i P_R) (k_4
\cos \beta+k_3 \sin \beta)] \tilde{\chi}^{0}_j A    \nn\\ 
&+&\bar{\tilde{\chi}}_i^{0} \tilde{\chi}^{+}
(V^T_{i\,3} P_L + V^\dag_{i\,3} \epsilon_i P_R)((k_2-k_3)U^C_{1\,2}+(k_4-k_1)\cos\beta)H^- +H.c. \nn \,, \label{potF}
\eea 
where $H_{n=1,2}=h,H$ are the two neutral scalars, $A$ the
pseudoscalar, $H^-$ the charged scalar, $\beta$ the mixing angle defined in \eq{hruot} and $U^R$ is the
mixing matrix related to the real  neutral 
components of the two doublets $h_1$ and $h_2$ given by
\bea
U^R&=&
\left(
\begin{array}{cc}
\cos\alpha &   - \sin \alpha  \\
 \sin \alpha & \cos \alpha  
\end{array}
\right)\,,
\eea
where $\alpha$ is the mixing angle defined in \eq{eqalfa}.

The analysis shows that the relic abundance is always at least one order of magnitude too small than the presently favorite abundance of dark matter in the Universe. This seems to be due to the lack of cancellations among different diagrams introduced by the arbitrariness in the Yukawa couplings that makes pair annihilation rates too large. Therefore, the lightest neutral exotic fermion can at most be a marginal component of dark matter.

\vskip1.5em
\section{The model solved} 
\label{sec:s}
The model has eleven parameters, 10 of which  are in principle free once the ground state has been identified with $v_W$. If we enforce the Veltman conditions---and thus make the one-loop quadratically divergent corrections vanish---we are left with eight parameters. These can be treaded for the masses of the 4 scalar and 4 fermion states. 
These can be varied and for each choice of them the S, T and U parameters computed and compared against the EW constraints. 

We vary the dimensionless parameters within one order of magnitude. In particular, we keep the $\lambda_i$  and the $\kappa_i$ between 1 and $4\pi$ (after which the the perturbative analysis may break down). Mass parameters $\mu$ and $\tilde \mu$ are varied between 100 and 300 GeV.

We find that for a large choice of the five remaining parameters the model is consistent with the EW precision measurements. For these choices, masses as large as 450 GeV are possible for the lightest neutral scalar Higgs boson. As its mass increases those of the neutral pseudoscalar  tends to favor lighter values so that there are solutions in which the lightest Higgs boson is the pseudoscalar. The lightest neutral fermion mass tends to increase together with the mass of the Higgs boson. 

\begin{figure}
\begin{center}
\includegraphics[width=4in]{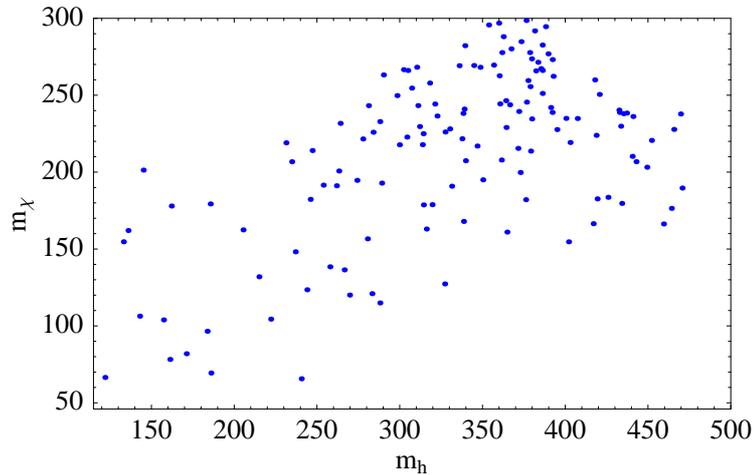}
\caption{\small  Distribution of values for the masses $m_\chi$ vs.\ $m_h$ for values of the parameters within 1$\sigma$ of EW precision masurements. \label{fig3}}
\end{center}
\end{figure}
\begin{figure}
\begin{center}
\includegraphics[width=4in]{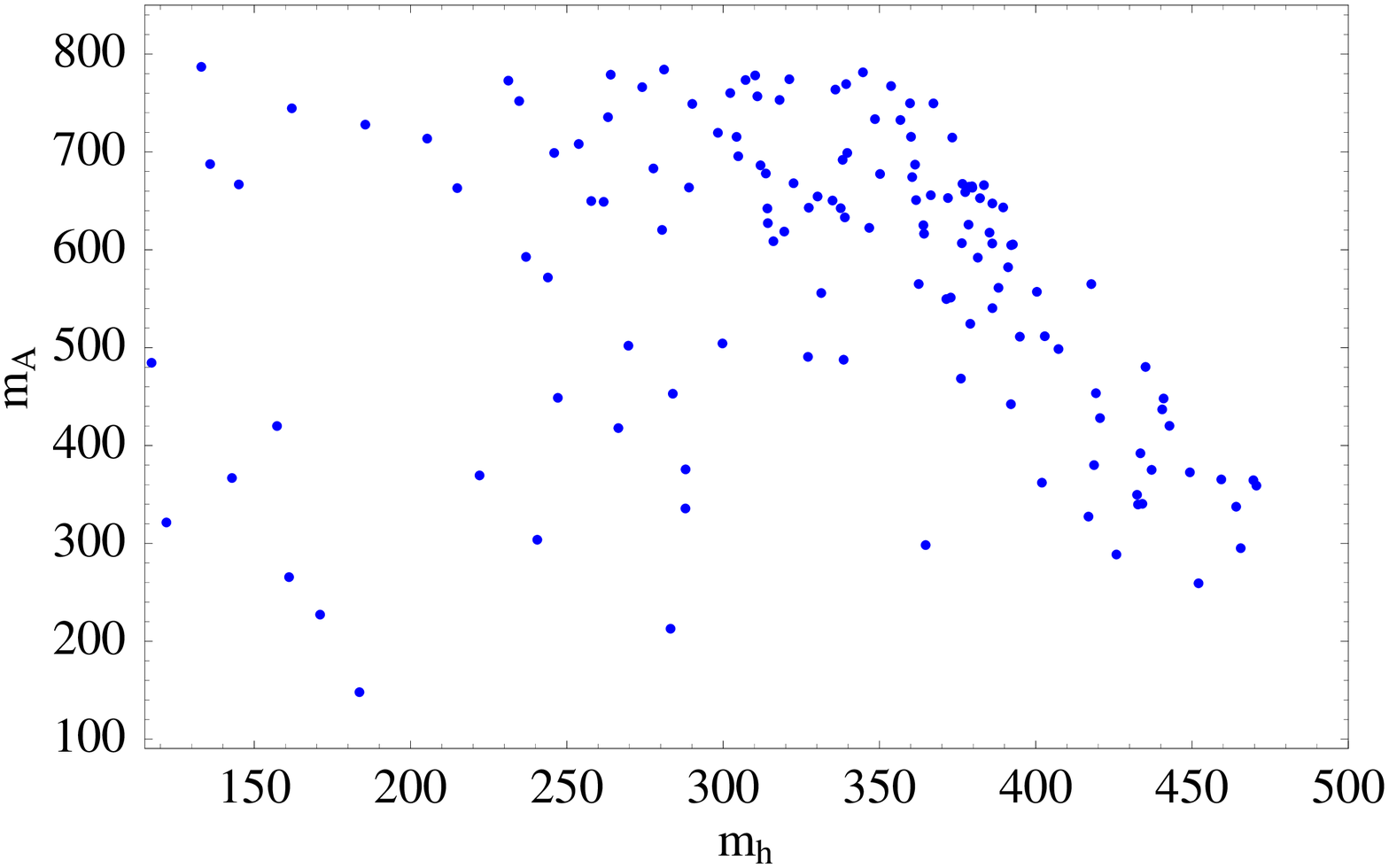}
\caption{\small  Distribution of values for the masses $m_A$ vs.\ $m_{h}$ for values of the parameters within 1$\sigma$ of EW precision masurements.  \label{fig4}}
\end{center}
\end{figure}
\begin{table}[ht]
\begin{center}
\caption{Representative values (among those used in the plots) of the eight parameters of the model, and mass spectrum of the most relevant states: scalar and pseudo-scalar bosons and lightest fermion, that satisfy the bounds from EW precision measerements.\label{tab1}}
\label{data}
\vspace{0.2cm}
\begin{tabular}{|cc|cccccc|ccc|}
\hline
 $\mu$ (GeV) & $\tilde \mu$ (GeV) & $k_1$ & $\lambda_1$& $\lambda_2$& $\lambda_3$& $\lambda_4$& $\lambda_5$ 
& $m_h$ (GeV) & $m_{A}$ (GeV) & $m_\chi$ (GeV) \cr
\hline
173& 287 &1.4 & 8.5& 4.6& 2.8 & $-8.5$ & $-5.6$ & 146 & 600  &  131 \cr
138 & 128  &1.6 & 6.3 & 6.4& 2.3 & $-10.6$ & $-2.9$ & 210 & 417  &  96  \cr
276 & 438 & 2.9 & 7.7 & 5.0 & 8.1 & $-7.1$ & $-8.5$ & 304  & 715  &  223  \cr
266 & 381 & 3.8 & 5.3 & 12.5 & 1.7 & $-7.0$ & $-3.5$ & 450  & 460 & 212  \cr
239 &180  &3.8 & 4.8& 11.4 & 7.3 & $-11.9$ & $-2.1$ & 470 & 360  &  190  \cr
\hline
\end{tabular}
\end{center}
\end{table}

Figure~\ref{fig3} shows some of the possible values we obtain for the Higgs boson and lightest neutral fermion masses for values of the parameters which satisfy within 1$\sigma$ the EW precision masurements. Figure~\ref{fig4} shows the distribution of the masses for the scalar and pseudoscalar states under the same conditions.

Our result may help in dispelling excessive surprise in not seeing a bantamweight  Higgs boson with $m_h$ just above the current LEP bound of 117 GeV and should encourage searches at the LHC for a Higgs boson substantially heavier than the current LEP bound---what we can call a welterweight at $m_h$ around 300 GeV or even a cruiserweight at  500 GeV. Such a scenario  has been pointed out recently in \cite{littlest} and \cite{flhiggs} in the framework of the little Higgs models\cite{LH}   and in \cite{barbieri1,barbieri2} in a two-Higgs extension of the SM.

\vskip1.5em
\section{Models with two Higgs bosons and no extra fermions} 
\label{sec:2}
Different possibilities of realizing a minimal extention of the scalar sector of the SM could have  a natural cut-off $\Lambda$ around few TeV while being  compatible with EW precision measurements have been discussed in the last year.  The authors of  \cite{barbieri1,barbieri2,Chacko}   have analyzed different realizations of the 2 Higgs doublets model  (2HDM)  and  have parametrized the fine tuning parameter in terms of the dependence of the mass  of the light Higgs boson  on the cut--off $\Lambda$.  In the Barbieri-Hall (BH) model~\cite{barbieri1} both doublets acquire a VEV, but the small mixing angle between them makes the light scalar coupling to the top quark quite small and  $\Lambda$ becames  proportional to the mass of the heavy neutral scalar. The mass of the heavy neutral scalar is then bounded by the requirement of satisfying the EW  precision measurements and this allows $\Lambda$ to reach more or less 2  TeV  when the  light Higgs boson has a mass $m_h=115$ GeV.  The twin doublets model~\cite{Chacko} is a particular version of the 2HDM in which only one doublet couples to the SM fermions. The symmetry of the model makes possible to improve the bound found in the BH model and to reach a cut-off between 3 and 4 TeV. Finally, the  inert doublet model (IDM)~\cite{barbieri2}  proposes a different picture. Instead of trying to justify through naturalness the existence of a light Higgs boson and  a cut-off of few TeV, it describes the possibility of having a heavy Higgs  while being still compatible with EW precision measurements. The cut-off of the model  turns out to be of few TeV (a value that would be natural even in the SM  context if the Higgs were heavy).  The new feature of the IDM is that the model may be compatible with the EW precision measurements even in the presence of a heavy Higgs boson. This is realized thanks to the contribution to t
he EW parameters that arises from the heavy new scalars. In general, in the different realizations of the 2HDM the T parameter receives a SM-like contribution and  a contribution that arises from loops involving the new scalars. These contributions are approximately given by \cite{Casas} 
\bea
T_a&=&-\frac{3}{16 \pi \cos \theta_W^2} (\cos^2( \alpha-\beta) \log \frac{m_h^2}{m_Z^2}+\sin^2( \alpha-\beta) \log \frac{m_H^2}{m_Z^2}) \nn\\
T_b&=&\frac{1}{4 \pi s_W^2 m_W^2}(\cos^2\alpha (m_{H^+}-m_h)^2+\sin^2\alpha (m_{H^+}-m_H)^2 + (m_{H^+}-m_A)^2\nn\\
&-&\cos^2\alpha (m_A-m_h)^2-\sin^2\alpha (m_A-m_H)^2 )\,,
\eea
where $\tan\beta= v_2/v_1$ with $v_i$ the  vev of $h_i$ and $\alpha$ the mixing angle between the two neutral scalars. If both the doublets acquire a VEV (BH,  twin and the just-so models)  $T_b$ is negligible because $m_A-m_{H^+}$ cannot be too large (for natural choice of the $\lambda_i$ parameter of the potential). On the contrary,  in the IDM $T_b$ may not be negligible and can balance the contribution  to $T_a$ arising from a heavy Higgs boson; in this way, the model predicts a heavy Higgs boson and a cut-off around $3$ TeV. In conclusion,  in all the version of 2HDM the cut-off can  be   around $5$ TeV but not much higher. 

Our approach is  different with respect to the models that present improved naturalness.  The cancellation of the Veltam condition fixes our cut-off at $10$ TeV and the requirements to be compatible with the EW precision measurements and to cover the most general neutral scalar spectrum  forces us to  include   at least a new fermion doublet.

\acknowledgments

 This work is
partially supported by  MIUR  and the RTN European  Program MRTN-CT-2004-503369.  F.~B.  is supported by a
MEC postdoctoral grant.


 \end{document}